\documentclass[prl,twocolumn]{revtex4}%
\usepackage{amsmath}
\usepackage{amsfonts}
\usepackage{amssymb}
\usepackage{graphicx}%
\begin{document}

%\preprint{}

\title{Liquid Crystal Tunable Metamaterial Perfect Absorber}

\author{David Shrekenhamer, Wen-Chen Chen, and Willie J. Padilla$^\star$}
\affiliation{Department of Physics, Boston College, 140 Commonwealth Ave., Chestnut Hill, MA 02467, USA}

\email{willie.padilla@bc.edu}

\begin{abstract}

We present experimental demonstration of electronically tunable metamaterial perfect absorbers in the terahertz regime. By incorporation of active liquid crystal into strategic locations within the metamaterial unit cell we are able to modify the absorption by 30 percent at 2.62 THz, as well as tune the resonant absorption over 4 percent in bandwidth. Numerical full-wave simulations match well to experiments and clarify the underlying mechanism, i.e. a simultaneous tuning of both the electric and magnetic response that allows for the preservation of the resonant absorption. These results show that the fundamental light interactions of surfaces can be dynamically controlled by all-electronic means and provide a path forward for realization of novel applications.

\end{abstract}

\maketitle
Designer materials consisting of sub-wavelength, periodic metallic inclusions may be fashioned to exhibit exotic electromagnetic effects that do not occur naturally, including negative index of refraction~\cite{Shelby01, Smith00}, invisibility cloaking~\cite{Schurig06}, and perfect lensing~\cite{Pendry00}. Metamaterial perfect absorbers (MPAs) have sparked significant interest since their first demonstration~\cite{Landy08} and hold great potential for use in applications ranging from thermal emitters~\cite{Liu11} and energy harvesting~\cite{Aydin11}, to sensors and spatial light modulators~\cite{Liu10}. The ability of sub-wavelength metamaterial unit cells to completely absorb incident electromagnetic energy have been demonstrated across much of the electromagnetic spectrum~\cite{Landy08,Tao08,Hao10}. One key design feature afforded by metamaterials is the ability to engineer materials with specified electric [$\epsilon(\omega)$] and magnetic [$\mu(\omega)$] response. This allows for matching the impedance $Z(\omega)=\sqrt{\mu(\omega)/\epsilon(\omega)}$ of the MPA to that of free space which results in zero reflection. The resonant nature of the MPA structure generates large electric field enhancement and thus completely dissipates incident energy; primarily by dielectric loss~\cite{Tao08b,Liu10}.

Although metamaterials have demonstrated many exotic effects, their resonant nature usually implies operation over a limited bandwidth. For various applications where bandwidth is not an issue, the ability to control the electromagnetic response is still highly desired. Thus in order to achieve practical devices, the creation of tunable structures is of paramount importance. In recent years frequency agile and tunable metamaterials have been demonstrated and various methods utilized to achieve amplitude and frequency modulation including photodoping~\cite{Chen08}, electronic charge injection~\cite{Chen06b} and temperature control~\cite{Driscoll08}. Liquid crystals are viable candidates for tunable / dynamic metamaterials with much unexplored potential. The voltage dependent birefringence exhibited by nematic liquid crystals is well known and has been exploited over the past several decades leading to the development of devices, most notably optical display technology. However, the birefringence in LCs extends well beyond the visible spectrum and has afforded a new outlet for electronically tunable metamaterial and photonic structures~\cite{Wilk09b,Khoo06,Werner07,Hu07,Gorkunov08,Minovich12}. Here we demonstrate that by combination of the variable properties of liquid crystals with metamaterials, an electronically tunable perfect absorber at terahertz (THz) frequencies is thus realizable.

\begin{figure}[!]
\begin{center}
\includegraphics[ width=3.5in,keepaspectratio=true
]%
{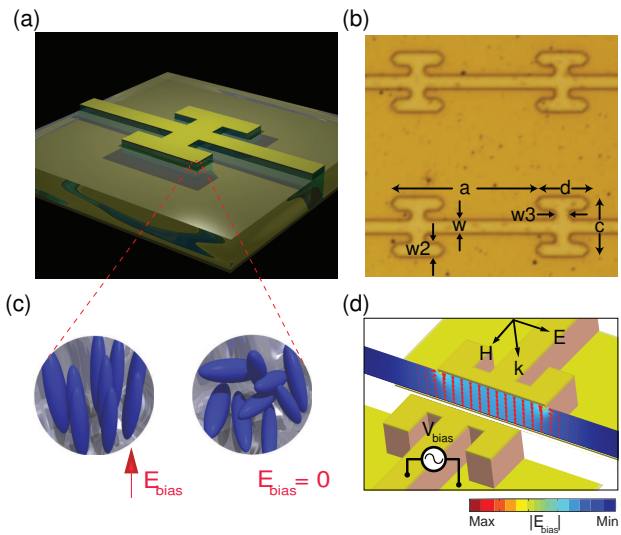}%
\caption{(Color online) Design and operational principle of the tunable metamaterial perfect absorber. (a)  Rendering of a single unit cell of the liquid crystal metamaterial perfect absorber. (b) Optical microscope image of a portion of the metamaterial array where the unit cell has the dimensions of \emph{a} = 50, \emph{c} = 20, \emph{d} = 16, \emph{w} = \emph{w}2 = 4.5, and \emph{w}3 = 5 in micrometers. (c) Depiction of the random alignment of liquid crystal in the unbiased case (right) and for an applied AC bias (left). (d) Simulation of the electric field vector and absolute value produced from an applied potential bias between the ERR and ground plane.}
\label{Fig1}%
\end{center}
\end{figure}

Metamaterial perfect absorbers typically  consist of two metallic layers spaced apart by virtue of a dielectric spacer. The top metal layer is geometrically patterned in order to strongly couple to a uniform incident electric field. By pairing the top layer with a metal ground plane, a mechanism for coupling to the magnetic component of light is created - see Fig.~\ref{Fig1}(a). Altering the geometry of the metallic pattern and dielectric thickness enables tuning of the effective material response parameters allowing for both impedance matching and strong absorption at nearly any desired frequency~\cite{Padilla10}. A schematic and optical microscope image of the device are shown in Figs.~\ref{Fig1}(a) and (b) with a single unit cell having the dimensions as labeled. Electric ring resonators (ERRs)~\cite{Schurig06b,Padilla07} were fabricated to form a square array with 50 $\mu$m lattice spacing. Each unit cell is connected to its neighbors via horizontal metallic wires (4.5 $\mu$m width) and the entire array is connected to bias pads lying at the perimeter of the device. A 200 nm Au/Ti continuous metal ground plane was E-beam deposited on top of a supporting silicon (Si) substrate. A 5.5 $\mu$m thick liquid polyimide (PI-5878G, HD Microsystems$^{TM}$) dielectric layer was spin coated on top. Ultraviolet (UV) photolithography is used to pattern photoresist which was used for final deposition of 200 nm Au/Ti to create the ERR layer. We also use the ERR structures to serve as a hard mask for inductively coupled plasma and reactive ion etching in order to remove all polyimide not directly underneath the metamaterial layer.

The liquid crystal 4'-n-pentyl-4-cyanobiphenyl (5CB) is deposited on top of the metamaterial array and completely fills in and encapsulates the polyimide / metal structure. 5CB possesses a nematic LC phase at room temperature with large birefringence ($n_e -  n_o =\Delta$n) at THz frequencies ranging between 0.11 to 0.21~\cite{Tsai03,Tsai04,Wilk09a,Vieweg10}, where the refractive index can be switched between its ordinary $n_{o}$ and extraordinary $n_{e}$ value in the presence of an electric field. A schematic shown in Fig.~\ref{Fig1}(a,c) illustrates the mechanism by which the LC is tuned. A potential is applied between the ERR and ground plane which orients the LC along field lines (see Fig.~\ref{Fig1}(d)). The polyimide is required for structural support however it plays a significantly more important role in our design. Most liquid crystals have large interactions with boundaries which inhibits any possible response to an applied electric field thus causing threshold phenomena -- an effect called the Freedericksz transition~\cite{Collings10}. Notably, our configuration permits LC near the surface of the polyimide to be \emph{orientated with electric field lines}. This also facilitates a smooth tuning of the refractive index as a function of applied electric field.

The frequency dependent reflection [$R(\omega)$] was characterized at an incident angle of 20 degrees from 2.0 to 3.5 THz using a Fourier-transform infrared spectrometer, liquid helium-cooled Si bolometer detector, and a germanium coated 6 $\mu$m mylar beamsplitter. The measured reflection spectra are normalized with respect to a gold mirror and we calculate the frequency dependent absorption as $A(\omega) = 1-R(\omega)$ since the transmitted intensity was zero due to the metal ground plane. Measurements were performed with the THz electric field perpendicular to the metal connecting wires, as depicted in Fig.~\ref{Fig1}(d). The LC molecules are aligned by applying a square-wave potential between the ERR metal layer and the ground plane at various modulation frequencies ($f_{mod}$). The square-wave is centered about zero and has peak-to-peak voltage equal to twice the peak bias voltage (V$_{bias}$). Use of a modulated bias prevents free carrier build-up at the electrode metal interface which can occur for DC applied potentials~\cite{Meier66,DeJeu78}.

\begin{figure}[!]
\begin{center}
\includegraphics[ width=2.0in,keepaspectratio=true
]%
{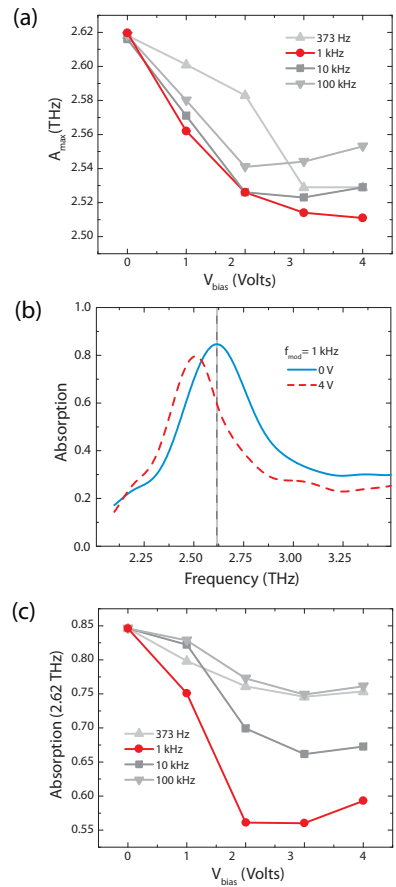}%
\caption{(Color online) Experimentally measured absorption of the metamaterial perfect absorber. (a) Frequency location of the absorption maximum ($A_{max}$) as a function of applied bias voltage (V$_{bias}$) for modulation frequency ($f_{mod}$) values of 373 Hz, 1 kHz, 10 kHz, and 100 kHz. (b) Frequency dependent absorption A($\omega$) for 0 V (blue solid curve) and 4 V (red dashed curve) at $f_{mod}$ = 1kHz, dashed line is centered at $A_{max}$(V$_{bias}$ = 0) = 2.62 THz. (c) The absorption value at 2.62 THz as a function of V$_{bias}$ for various modulation frequencies.}
\label{Fig2}%
\end{center}
\end{figure}

The absorption was characterized for a number of different bias values and modulation frequencies. In Fig.~\ref{Fig2}(a) we show the frequency location of the absorption maximum ($A_{max}$) as a function of V$_{bias}$ for modulation frequencies of 373 Hz, 1 kHz, 10 kHz, and 100 kHz. The general trend is that as we increase the applied voltage the metamaterial absorption shifts to lower frequencies. For $f_{mod}$ = 373 Hz and 1 kHz the change is monotonic for increasing potential, but we find deviations from this for 10 kHz and 100 kHz. As can be observed in Fig.~\ref{Fig2}(a) the greatest frequency shift occurs for $f_{mod}$ = 1 kHz - in accord with prior investigations with 5CB~\cite{Tsai04,Wilk09a,Vieweg10}. In Fig.~\ref{Fig2}(b) we plot the frequency dependent absorption A($\omega$) for 0 V (blue solid curve) and 4 V (red dashed curve) at $f_{mod}$ = 1 kHz. With no applied bias we achieve a reasonable absorption of 85\% at 2.62 THz, a full width half max (FWHM) of 600 GHz and the spectrum is otherwise featureless. At V$_{bias}$ = 4 V the resonant absorptive feature shifts to 2.5 THz, lowers to a peak value of 80\% and narrows slightly with a FWHM of 420 GHz. This represents a shift in the peak of the absorption by 4.6\% in frequency.

In many applications one may desire amplitude modulation only over a narrow band. As an example we consider operating at a fixed frequency of $\omega_0$ = 2.62 THz, i.e. the peak absorption of the unbiased case plotted in Fig.~\ref{Fig2}(b). In Fig.~\ref{Fig2}(c) we plot $A(\omega_0)$ as a function of V$_{bias}$ for various modulation frequencies. Generally we observe that the absorption level drops as a function of increasing voltage bias for all modulation frequencies investigated, which seems to saturate near 3-4 volts of applied bias. The greatest change in $A(\omega_0)$ occurs for 1 kHz bias modulation, as shown by the red curve in Fig.~\ref{Fig2}(c) -- in accord with the results presented in Fig.~\ref{Fig2}(a). The LC 5CB thus provides an all electronic means of both frequency and amplitude tuning of the absorption peak of metamaterial perfect absorbers and here we realize an amplitude tuning of over 30\% at $\omega$ = 2.62 THz.

In order to clarify the mechanism underlying operation of the tunable absorber, we perform full wave 3D electromagnetic simulations. The Au/Ti metal layers were modeled as a lossy metal with a frequency independent conductivity of $\sigma=4.56\times10^7$ S/m, and the polyimide layer with a relative permittivity of $\tilde{\varepsilon}_{poly}$ = $\varepsilon_{1}$ +  i$\varepsilon_{2}$ = 2.88 + i$0.09$~\cite{Tao08c}. In accord with prior work~\cite{Tsai03,Tsai04,Wilk09a,Vieweg10}, we model the complex refractive index of 5CB, (with zero applied bias), as a lossy dielectric with $\tilde{n}_{5CB}$  =  $n_{5CB}$ +  i$\kappa_{5CB}$ = 1.82 + i$0.14$. As mentioned, the LC encapsulates the metamaterial array and thus we model a 2 $\mu$m thick layer on top of the ERR. We assume that any LC not lying in-between the ERR and ground plane is unaltered by the applied bias (see Fig.~\ref{Fig1}(d)). The THz birefringent properties of 5CB have been characterized as an increase in the real part of the refractive index ($n_{5CB}$), (for increasing applied bias), between the ordinary and extraordinary states~\cite{Tsai04,Wilk09a}. However, it has been demonstrated that there is little difference in the imaginary component ($\kappa_{5CB}$) above 1.2 THz~\cite{Tsai03}. In simulation, we thus modify only $n_{5CB}$ as a function of V$_{bias}$ and keep the imaginary refractive index constant at a value of $\kappa_{5CB}$ = $0.14$.

\begin{figure}[!]
\begin{center}
\includegraphics[ width=3.5in,keepaspectratio=true
]%
{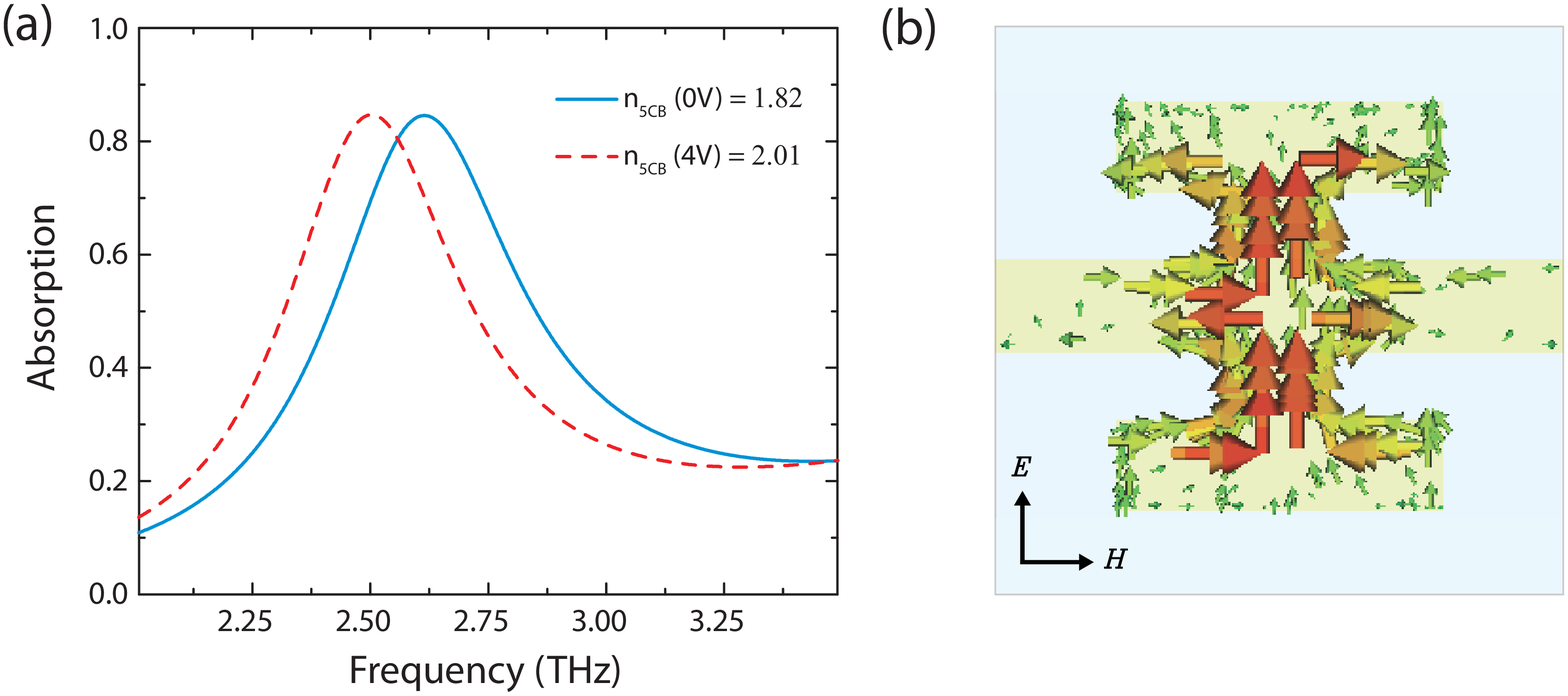}%
\caption{(Color online) Numerical simulations of the metamaterial perfect absorber. (a) THz absorption simulated for 0 V (blue solid curve) and 4 V (red dashed curve) at $f_{mod}$ = 1kHz. (b) Simulated current density in the ERR for the case of the unbiased absorption maximum $A_{max}$(V$_{bias}$ = 0) = 2.62 THz.}
\label{Fig3}%
\end{center}
\end{figure}

Results from the computational investigation are presented in Fig.~\ref{Fig3}. As previously mentioned increasing the applied potential results in an increase in $n_{5CB}$. We thus monotonically increase the real part of the refractive index which results in a redshift of the absorption peak frequency (see Fig.~\ref{Fig3}(a)). At 4 V the peak absorption occurs at 2.51 THz and in the numerical model we determine $\tilde{n}_{5CB}$ (V$_{bias}$ = 4)  = 2.01 + i$0.14$. As can be observed we find good agreement with experimental results (see Fig.~\ref{Fig2}), although simulation indicates that the value of the peak absorption at 4 V applied bias is not significantly altered from the unbiased state. Notably we find that the change in refractive index determined by simulation is $\Delta$n = 0.19 -- consistent with values determined by other works~\cite{Tsai03,Tsai04,Wilk09a,Vieweg10}. At a frequency of 2.62 THz, simulation predicts a change in $A$ of 15\% between zero and 4 V of applied bias, as shown in Fig.~\ref{Fig3}(a). In contrast experimental results yield a 30 percent change in absorption at the same frequency.

\begin{figure}[!]
\begin{center}
\includegraphics[ width=3.3in,keepaspectratio=true
]%
{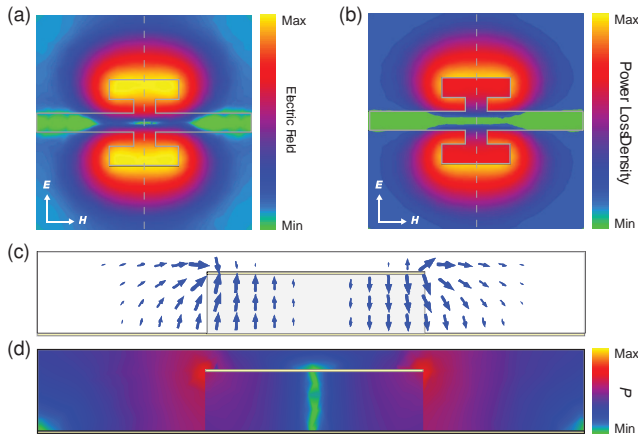}%
\caption{(Color online) Resonant terahertz fields of the metamaterial perfect absorber. (a) Simulated electric field magnitude and (b) power loss density ($P$) shown at a plane centered between the two metallizations. (c,d) Terahertz electric field vector plot (c) and power loss density (d) shown for a cross sectional cut (gray dashed line in (a,b)).}
\label{Fig4}%
\end{center}
\end{figure}

It is instructive to examine the particular mode exhibited by our device at the maximum of the absorption, i.e. 2.62 THz. This can be explored by observation of the the surface current density and magnitude of the THz electric field (plotted in a plane centered between the two metallizations), as shown in Figs.~\ref{Fig3}(b) and~\ref{Fig4}(a). We find that the surface current density is similar to that found in prior investigations~\cite{Schurig06b} and, as expected, the THz electric field is primarily localized underneath the ERR - in the same vicinity as the electric field provided by the bias shown in Figure 1c. In contrast, the power loss density shown in Fig.~\ref{Fig4}(b), reaches its strongest values just outside the ERR at the polyimide / LC interface. (The magnitude of the electric field is also plotted as a vector field in Fig.~\ref{Fig4}(c), and the power loss density in Fig.~\ref{Fig4}(d); both in cross section.) The form of the metamaterial absorptive feature strongly depends on the value of the complex dielectric constant that the local terahertz electric field experiences. In particular the resonant frequency is set by the real part of the dielectric function, whereas the width of the absorption is determined by dielectric loss~\cite{Tao08b,Liu10}. Thus future designs can achieve greater frequency tuning of the absorption peak by altering the geometry such that the LC lies directly underneath the ERR, where the applied bias is greatest, as shown in Fig.~\ref{Fig1}(d). An alternative approach would be utilization of LC polymers~\cite{Rutz06}, which could then act both as a supporting structure, and bias tunable dielectric.

\begin{figure}[!]
\begin{center}
\includegraphics[ width=2.5in,keepaspectratio=true
]%
{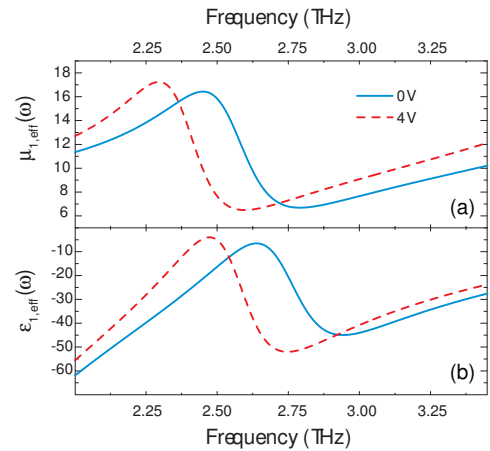}%
\caption{(Color online) Effective optical constants of metamaterial perfect absorber. (a) Real part of effective permeability ($\mu_{1,eff}$), and (b) real part of effective permittivity ($\epsilon_{1,eff}$), determined from inversion of the simulated scattering parameters for 0 V (blue solid curves) and 4 V (red dashed curves).}
\label{Fig5}%
\end{center}
\end{figure}

We note that in order to frequency tune a metamaterial in which both the electric and magnetic properties have been designed, it is desirable to adjust them identically to preserve the electromagnetic response. Some dynamic magneto-dielectric metamaterials may require sophisticated tuning mechanisms in order to maintain their properties~\cite{Mosallaei07}. In contrast, a salient feature of this particular perfect absorber design, is the ability to simultaneously adjust $\epsilon(\omega)$ and $\mu(\omega)$ by simply altering the dielectric properties of the dielectric spacing layer, as demonstrated here with liquid crystal. This can be verified by plotting the extracted~\cite{Liu10} material parameters (assuming a thickness of 7.7 $\mu$m equal to its physical dimensions~\cite{Chen12}) for the perfect absorber obtained from simulations utilizing a frequency dependent Drude model (plasma frequency $\omega$$_p$ = 2$\pi$$\times$2175 THz and collision frequency $\omega$$_c$ = 2$\pi$$\times$6.5 THz). As revealed in Fig.~\ref{Fig5}, both the permittivity and permeability shift with little change in their shape for all applied biases investigated.

The demonstration of the dynamic control of electromagnetic waves at surfaces presented here provides a new path forward construction of exotic devices. The formation of pixelated arrays of metamaterial LCs would provide a means for spatial light modulation (SLM)~\cite{Chan09}, (including phase and amplitude), and focal plane array imagers. As both metmaterials and liquid crystals have been shown to operate across much of the electromagnetic spectrum this would enable an innovative substitute for digital micromirror devices (DMDs), which are spectrally featureless and have set frequency limits~\cite{Duarte08}. Modern imaging research that could benefit from SLMs, such as compressive sensing~\cite{Donoho06}, Hadamard imaging~\cite{Pratt69} and adaptive coded aperture imaging~\cite{Slinger07}, would then be enabled at EM bands where conventional DMDs do not operate. The fabrication of focal plane arrays consisting of the designs demonstrated here could serve as an excellent candidate for detector pixels when implemented into bolometric, rectification, pyroelectric and piezoelectric configurations.

In conclusion, we have demonstrated all-electronic control of liquid crystal metamaterial perfect absorbers. We achieve a 30 percent amplitude tuning of the absorption and realize a frequency tunability greater than 4 percent. The scalability of both liquid crystal properties and metamaterial perfect absorbers suggests our design can be extended to both higher and lower frequencies. The prospect of electronically controlled metamaterial perfect absorbers have implications in numerous scientific and technological areas rich in applications, particularly in sensing, imaging, energy harvesting and dynamic scene projectors.

We acknowledge support from the Department of Energy under grant number DE-SC0005240 and the Office of Naval Research under U.S. Navy Contract Nos. N00014-07-1-0819, N00014-11-1-0864.

\end{document}